\documentclass[%
 amsmath,amssymb,
preprint,%
 reprint,%
]{revtex4-1}

\usepackage{graphicx}
\usepackage{dcolumn}
\usepackage{bm}

\usepackage[utf8]{inputenc}
\usepackage[T1]{fontenc}
\usepackage{mathptmx}
\usepackage{etoolbox}

\makeatletter
\def\@email#1#2{%
 \endgroup
 \patchcmd{\titleblock@produce}
  {\frontmatter@RRAPformat}
  {\frontmatter@RRAPformat{\produce@RRAP{*#1\href{mailto:#2}{#2}}}\frontmatter@RRAPformat}
  {}{}
}%
\makeatother
\begin{document}

\preprint{Cooke, Jerolmack, and Park}

\title[Sample title]{Mesoscale structure of the atmospheric boundary layer across a natural roughness transition}
\author{Justin P. Cooke}
\author{George I. Park}%
    \homepage{Corresponding Author Email: gipark@seas.upenn.edu.}
\affiliation{ 
Mechanical Engineering and Applied Mechanics, \\ University of Pennsylvania, Philadelphia, PA 19104
}%

\author{Douglas J. Jerolmack}%
    \altaffiliation[Also at ]{Mechanical Engineering and Applied Mechanics, University of Pennsylvania.}
\affiliation{Earth and Environmental Science, University of Pennsylvania, Philadelphia, PA 19104 \\}

\date{\today}

\begin{abstract}
The structure and intensity of turbulence in the atmospheric boundary layer (ABL) drives fluxes of sediment, contaminants, heat, moisture and CO$_2$ at the Earth's surface.  
Where ABL flows encounter changes in roughness -- such as cities, wind farms, forest canopies and landforms -- a new mesoscopic flow scale is introduced: the internal boundary layer (IBL), which represents a near-bed region of transient flow adjustment that develops over kilometers.  
This important scale lies within a gap in present observational capabilities of ABL flows, and simplified models fail to capture the sensitive dependence of turbulence on roughness geometry.
Here we use {large-eddy simulations}, run over high-resolution topographic data and validated against field observations, to examine the structure of the ABL across a natural roughness transition: the emergent sand dunes at White Sands National Park. 
We observe that development of the IBL is triggered by the abrupt transition from smooth playa surface to dunes; however, continuous changes in the size and spacing of dunes over several kilometers influence the downwind patterns of boundary stress and near-bed turbulence. Coherent flow structures grow and merge over the entire $\sim$10-km distance of the dune field, and modulate the influence of large-scale atmospheric turbulence on the bed. Simulated boundary stresses in the developing IBL explain the observed downwind decrease in dune migration, demonstrating a mesoscale coupling between flow and form that governs landscape dynamics. More broadly, our findings demonstrate the importance of resolving both turbulence and realistic roughness for understanding fluid-boundary interactions in environmental flows.
\end{abstract}

\maketitle

\begin{quotation}
\noindent \textbf{Significance Statement:} 
\\

We live within the Atmospheric Boundary Layer (ABL), where air flow feels the friction of the planet's surface, producing turbulence. 
When ABLs encounter changes in roughness -- from sea to land, or rural to city -- a near-surface region with distinct turbulence characteristics develops. 
The structure within influences transport of heat, water, and substances like wildfire smoke; yet, data in this region are sparse, with existing models unable to capture important behaviors. 
We use advanced computational tools to identify patterns in the turbulence structure over a roughness transition in a sand dune field, validating our model against rare field observations. 
Our results explain feedback between flow and topography that influence this landscape, revealing behaviors that may be common across natural roughness transitions. 
\end{quotation}

\section*{\label{sec:level1} Introduction}

Whenever turbulent flows impinge on a surface, a boundary layer develops -- a thin region near the surface where the {shear stress} of the boundary is felt by the flow, characterized by a steep velocity gradient \cite{sreenivasan1989turbulent,pope}. 
The stress exerted by the flow on the boundary depends sensitively on the geometry of boundary roughness itself; thus there is a feedback between flow and form \cite{chung2021predicting,clauser}. 
Even for the canonical case of uniform sand grains glued to a pipe, surface drag is a complex and non-monotonic function of the {roughness} Reynolds number $k_s^+ \equiv u_\tau k_s/\nu$, where $u_\tau$ is the friction velocity, $\nu$ is the kinematic viscosity of the fluid and $k_s$ is the equivalent sandgrain roughness, a hydraulic length-scale defined by Nikuradse~\cite{nikuradse1933,chung2021predicting}. 
Seemingly {minor} changes to the geometry and spacing of roughness elements can produce drastically different drag effects~\cite{chung2021predicting}, that cannot be {adequately resolved} with simple turbulence closures, such {as those used in} Reynolds Averaged Navier-Stokes (RANS) models~\cite{altland2021modeling}. 
A next level of complexity involves spatial changes in roughness, which trigger the growth of a near-bed region of transient flow adjustment called an Internal Boundary Layer (IBL)~\cite{elliott1958growth,antonia1971response,antonia1972response,li2019recovery}.
There are many applications that involve turbulent flows encountering complex and spatially-varying roughness, and where it is critical to know the boundary stress -- from wind turbines, to aircraft wings, to marine infrastructure \cite{chung2021predicting,kuwata2023scaling}. 
 
We live within the Atmospheric Boundary Layer (ABL), the roughly 1-km thick ({$\delta_{ABL} \sim 10^3$ m}) surface layer where flow interacts with topography and human-built structures \cite{bouzeid2020persistent, gul2022experimental,abedi2021numerical}. 
Due to the large length scales and highly heterogeneous roughness involved, ABL flows are highly turbulent {with friction Reynolds numbers $Re_{\tau} \equiv u_\tau \delta_{ABL}/\nu \sim O(10^6-10^7)$}. 
Step changes in roughness occur in many places, and have important consequences. 
Public health in cities is influenced by the interaction of ABL flows with buildings, and the associated fluxes of soot and heat carried by those flows \cite{li2013synergistic,manoli2019magnitude}. 
Effluxes of CO$_2$ and dust from landscapes -- critical components of the climate system \cite{baldocchi2020eddy, kok2023mineral}
 -- are also influenced by roughness transitions; sea to land, agricultural field to forest, or flow encountering mountains and dunes \cite{reth2005co2,baldocchi2000measuring}. 
In their recent review paper, Bou-Zeid and colleagues \cite{bouzeid2020persistent} categorized ABL flows based on the complexity of roughness, and noted that the category of irregular, heterogeneous roughness ``remains understudied, and a formal approach to understand the complex flow patterns over such surfaces and their regionally-averaged characteristics is critically lagging''. 
Resolving the structure and dynamics of the IBL, which develops from spatial changes in roughness, is hard to do with observations alone. 
Collecting time and height resolved flow data is expensive and technically challenging, making spatial coverage sparse \cite{abedi2021numerical,bell2020confronting}. 
Many field studies rely on at-a-point time series data, using Taylor's hypothesis to shift observations to the spatial domain \cite{hutchins2007evidence}; this approach is questionable for the highly non-uniform flow conditions of IBLs, {as many downstream stations are required to elucidate spatial variation due to its growth} \cite{li2019recovery,li2021experimental,hanson2016development,gul2022experimental}. 
More, natural ABL flows are non-stationary and often influenced by buoyancy effects \cite{cheng2005pathology, gunn2021circadian}, making it difficult to isolate the influence of roughness. 
Existing analytical formulas for describing the development of an IBL do a reasonable job of predicting the time-averaged scale of IBL thickness \cite{wood1982internal,elliott1958growth,pendergrass1984dispersion,savelyev2001notes,townsend1965response,panofsky1973tower,panofsky1984atmospheric}. 
Schemes for computing the velocity profile within the IBL, however -- and the resultant boundary stress -- may be limited in their applicability to irregular, heterogeneous roughness \cite{bouzeid2020persistent}. 
Due to these challenges, Large-Eddy Simulation (LES) has emerged as the tool of choice for examining spatially-resolved mesoscopic dynamics of the ABL \cite{anderson2020large}. 
LES parameterizes smaller-scale (inertial) turbulence while resolving larger eddies, {reducing computational cost,} which {allows for the exploration of much higher Reynolds numbers} than direct numerical simulation \cite{bose2018wall,choi2012grid,yang2021grid}. 
Using LES for ABL simulations over natural topography, however, can be prohibitively expensive computationally \cite{anderson2020large}; this means that compromises must sometimes be made. 
For example, the Immersed Boundary Method (IBM) is often chosen to approximate the solid surface \cite{verzicco2023immersed,anderson2020large}; however, this approach {is limited in its ability to confine grid refinement to near-wall regions, resulting in overly fine meshes away from the wall, and incurring high computational cost for flows at very-high $Re_\tau$~\cite{verzicco2023immersed}. 
Further, IBM} does not directly extract the boundary stress from the wall, requiring interpolation, compared to (more computationally-demanding) surface-conforming meshes \cite{verzicco2023immersed,constant2021improved}.
In addition, coarse computational meshes may limit flow resolution in the critical near-bed region, and field data are often insufficient to validate model choices.

Here we use LES to examine the flow across a natural roughness transition at White Sands National Park: sand dunes that emerge abruptly from a smooth playa. 
While previous LES studies have examined the variation in surface winds over individual dunes and dune clusters \cite{anderson2014numerical,wang2019turbulence}, they did not examine an entire dune field. 
We choose White Sands because it has been proposed that IBL development, triggered by the dunes, drives a downwind change in boundary stress that controls the migration of the dunes themselves -- with knock-on consequences for the hydrology and ecology of the region \cite{jerolmack2012internal, gunn2020macro, reitz2010barchan, lee2019imprint}. 
White Sands is also a significant source of dust in the region \cite{rea2020tracing,scheidt2010determinnig}. 
Our previously collected lidar velocimetry has shown how diurnal forcing drives a daily rhythm of near-surface flow due to buoyancy effects \cite{gunn2021circadian}; and our analysis of high-resolution topographic data has documented a spatial trend in dune migration that is consistent with IBL development \cite{jerolmack2012internal, gunn2020macro}. 
We perform a numerical experiment in which a steady and neutrally-buoyant flow is introduced over White Sands topography (Fig. 1\textit{A}). 
This allows us to isolate the effect of spatially-varying roughness on IBL development over $\sim~ 6$ km, and to elucidate the nature of the hypothesized coupling between topography and boundary stress at the mesoscopic scale.
Our simulation uses a surface-conforming mesh to directly resolve topography, and a highly-resolved and large numerical domain to capture a wide range of turbulence scales. 
Steady flow simulations reproduce the observed time-averaged velocity profile of wind over the smooth playa surface. 
Simulated IBL thickening downwind of the roughness transition is consistent with classic scaling models; however, changes {to the flow} are not as smooth or abrupt as simplified treatments would suggest. 
The flow responds continuously to changes in the spacing and geometry of dunes throughout the field, and the resulting boundary stress, $\tau_b$, is inconsistent with simplified composite schemes. Nevertheless, the vertical structure of turbulence within the developing IBL exhibits a robust self-similar structure.
Simulated patterns of boundary stresses explain the downwind slow-down of dunes previously reported \cite{jerolmack2012internal, gunn2020macro}, and reveal how meso-scale turbulent structures grow across the entire 6-km dune field. 
Our study shows the profound influence of complex roughness geometry on boundary stress, and how this can be isolated from other complicating factors in ABL flows using LES. 
These findings have relevance for other roughness transitions in natural ABL flows. 

\section*{Results}

\begin{figure*}[ht!]
   \centering
    \includegraphics[width=1\linewidth]{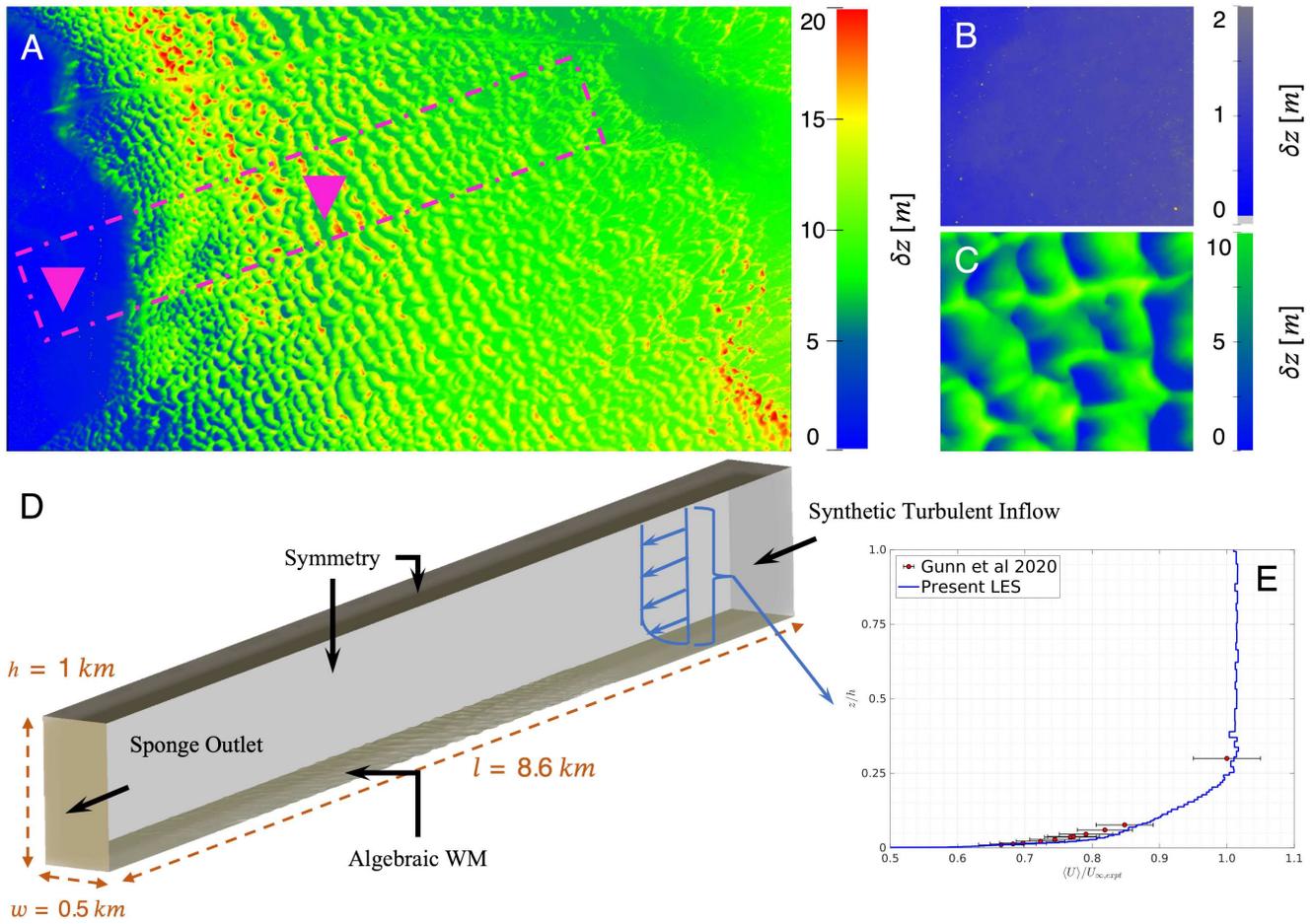}
    \caption{Topography and numerical setup. 
    (\textit{A}) Topographic lidar scan of White Sands National Park Dune Field, New Mexico, annotated with lidar wind velocity measurement locations (magenta triangles) and simulation domain (magenta dot-dashed box). 
    (\textit{B,C}) Topographic scans representative of the roughness levels at the velocity measurement locations upstream (\textit{B}) and downstream (\textit{C}) of the roughness transition. 
    (\textit{D}) Setup of the computational domain, with a synthetic turbulent inflow at the inlet, a sponge outflow at the outlet, symmetry conditions on the side and top walls, and an algebraic wall-model is applied to the bottom wall, which is synthesized from the topographic scan of the dune field. 
    Dimensions of the domain are given (orange). 
    (\textit{E}) The mean streamwise velocity profile from the LES (blue line) at the upstream field observation location is compared to the time-averaged experimental data with 5\% measurement uncertainty (black, red-filled circles). The data are normalized with the freestream velocity of the experiment, $U_{\infty,expt}$, and the wall elevation is normalized by the height of the domain, $h$.}
    \label{fig:fone}
\end{figure*}

\subsection*{Field data}
\label{sec:fielddata}
The field data used for this study have been extensively described elsewhere, and so are only briefly outlined here. 
Open-source topographic data \cite{usgs} for White Sands were gridded at 1 meter spatial ($x-y$) resolution, and have a vertical ($z$) resolution of $\sim 0.1$ m \cite{gunn2020macro}. 
Topography begins ($x=0$) on a smooth playa surface known as the Alkali flat, and rises relatively rapidly ($x = 1.8$ km) into a region we consider to be the start of the roughness transition where large transverse dunes abruptly emerge ($x \approx 1.9$ km). 
Within a kilometer of this transition, transverse dunes break up into isolated and heterogeneous barchan dunes, whose migration rate and amplitude decline gradually over several kilometers until the dunes are immobilized by vegetation \cite{jerolmack2012internal, reitz2010barchan, lee2019imprint}.
Flow velocity data come from the Field Aeolian Transport Events (FATE) campaign, collected from a fixed position on the smooth playa upwind of the dune field, with additional data collected on the stoss side of a dune downstream \cite{gunn2021circadian} (Fig. 1\textit{A, B,} and \textit{C}). 
A Campbell Scientific ZephIR 300 wind lidar velocimeter collected vertical velocity profiles every 17 seconds over approximately 70 days during the spring windy season of 2017 at White Sands, with a vertical resolution of 10 log-spaced bins from $z$ = 10 m to $z$ = 300 m above the surface and an additional point at $z$ = 36 m. 
Due to stratification effects, night-time winds produce a nocturnal jet aloft that skims over a surface layer of cool stagnant air \cite{gunn2021circadian}; as a result, boundary roughness effects and sand transport are suppressed at night. We average over all daytime measurements, {a twelve hour window from 06:00 to 18:00 local time,} to produce a time-averaged daytime velocity profile, that is used to validate the inflow conditions for our LES simulations. 

\subsection*{Numerical Setup and Validation}
\label{sec:setup}
We perform wall-modeled LES (WMLES) using the \textit{Charles} code from Cascade Technologies (Cadence Design Systems) \cite{gorle2023investigation,goc2021large,hwang2022large,cooke2023numerical,lozanoduran2022performance}, which is an unstructured grid, body-fitted finite-volume LES flow solver (\textit{Materials and Methods}). 
We numerically analyze a neutrally-buoyant and steady atmospheric boundary layer flow over an 8.6- by 0.5-km domain of the White Sands topographic data (Fig. 1\textit{D}), oriented in the direction of dominant winds and dune migration ($\sim$ 15 degrees N of E). 
The length of the domain is chosen to capture the mesoscopic scale of IBL development, and the width is chosen to be much larger than an individual dune. 
Prior field studies at White Sands have estimated the ABL thickness, $\delta_{ABL}$, to fluctuate daily between $O(10^2-10^3m)$~\cite{gunn2020macro,gunn2021circadian}. 
The height of the domain is chosen to be {$h = 1000$ m} to capture the upper-limit of this estimate, {and $\delta_{ABL}$ is chosen to be at $z$ = 300 m, the height of the highest velocity measurement.}
Simulations presented have a resolution of $0.75$ m closest to the boundary, and $28$ m in the outer-region of the flow (\textit{Materials and Methods}). 
A sensitivity analysis confirmed that our results are insensitive to domain size and resolution choices (\textit{SI Appendix}, Figs. S3-S5).
To minimize numerical effects, we deploy a numerical sponge outflow condition~\cite{mani2012analysis,bodony2006analysis} and at the sidewalls we use a symmetry boundary condition. 
{A synthetic turbulent inflow generation based on digital filter techniques~\cite{klein2003digital} is implemented at the inlet, and the numerical domain of the smooth Alkali Flat is extended to ensure that the inflow achieves a uniform condition before encountering the roughness transition (\textit{SI Appendix}).}
Simulations are run for over 1,000 large-eddy turnover times, $T \equiv \delta_{ABL}/U_\infty$ (\textit{SI Appendix}), to achieve steady conditions and allow convergence of various time-averaged flow quantities presented below. 

Streamwise, spanwise, and wall-normal coordinates within the domain correspond to $x$, $y$, and $z$, respectively, and have instantaneous velocity components $U$, $V$, and $W$. Using the Reynolds decomposition, $U = \langle U\rangle + u$, instantaneous velocities may be decomposed into time averaged (bracketed) and fluctuating (lower-case) quantities~\cite{pope}. 
Averaging over the spanwise direction is represented with barred notation, \textit{i.e.}, $\Bar{\cdot}$.
Quantities with normalization based on friction velocity and kinematic viscosity are signified by $\cdot^+$. 
We denote streamwise location in relation to the roughness transition with $\hat{x} = x - x_0$, $x_0$ being the approximate location of the start of the dune field.
{We first check that the inflow conditions have reached a fully developed state, matching a canonical zero-pressure-gradient turbulent boundary layer, by comparing skin-friction coefficient to the momentum thickness Reynolds number, $Re_\theta$, leading up to the transition (\textit{SI Appendix}, Fig. S1).}
We {then} test the validity of our model choices by comparing the time-averaged horizontal velocity profiles, $\langle U \rangle$, of our simulation and the observations of FATE on the Alkali Flat. 
The two agree over the entire measured elevation range to within 5\% (Fig. 1\textit{E}). 
This is remarkable, considering that the field data average over non-stationary forcing and buoyancy effects that are not modeled in the simulation. This agreement indicates that treating the ABL flow at White Sands as steady and neutrally buoyant is appropriate for describing time-averaged behavior. 

\subsection*{Characterization of the Internal Boundary Layer}
\label{sec:ibl}
With the validation in hand, we now use the simulations as a numerical experiment to examine how the IBL would develop due to the roughness transition under a steady and neutrally-buoyant flow. 
The qualitative flow behavior we observe is consistent with expectations and previous work \cite{gul2022experimental}. 
At the location where dunes emerge ($\hat{x} = 0$), there is an increase in turbulence and a shift of the high-velocity region farther from the bed (Fig. 2\textit{A}). 
We observe a gradual thickening of the perturbed flow region indicating a developing IBL. 
To characterize IBL growth we implement the method of Li \textit{et al.}~\cite{li2021experimental}, which uses streamwise variations in the streamwise turbulence intensity $\langle uu\rangle$, to define the thickness of the IBL, $\delta_{IBL}$:

\begin{equation}
    \Delta \Bigg[\frac{\langle uu\rangle}{U^2_\infty}\Bigg] \Big{/} \Delta \Bigg[\log_{10}(\frac{\hat{x}}{\delta_{ABL}})\Bigg] \rightarrow 0.
    \label{eqn:delta_ibl}
\end{equation}

\noindent This expression defines that, for successive downwind locations, the wall-normal height in which {the normalized value of $\langle uu \rangle$ divided by the normalized distance between stations} tends towards zero is equal to $\delta_{IBL}$ at the upstream streamwise station (\textit{SI Appendix}). 
For our simulations we choose a threshold value of $10^{-4}$ to represent convergence toward zero in Equation~\ref{eqn:delta_ibl} (\textit{SI Appendix}, Fig. S8). 
We verified that our results are insensitive to the choice of method for defining the IBL (\textit{SI Appendix}, Fig. S10 and Tables S2 and S3).
Ten velocity probing stations, logarithmically-spaced in the downwind flow direction from $\hat{x} = 50$ m to $\hat{x} = 5750$ m (\textit{SI Appendix}, Fig. S2), were placed to capture the growth of the IBL (Fig. 2\textit{B}). 
It is common to characterize a roughness transition as an abrupt change in the roughness parameter, $z_{01} \rightarrow z_{02}$, and to model IBL growth downwind of this transition as a smooth and monotonic function \cite{elliott1958growth, wood1982internal}.  
Gunn \textit{et al.}~\cite{gunn2020macro} identified the characteristic roughness parameters for the Alkali flat and the dune field to be $z_{01}= 10^{-4}$ m and $z_{02}= 10^{-1}$ m, respectively. 
Following previous work \cite{gul2022experimental,li2019recovery} we fit simulation data with a power law relation, $\delta_{IBL}/z_{02} = a_0(x/z_{02})^{b_0}$, and determine $a_0 = 0.29$ and $b_0 = 0.71$. 
This observed thickening of the IBL, induced by the smooth $\rightarrow$ rough transition associated with dunes, is consistent with classic scaling models [\textit{SI Appendix}, Fig. S9~\cite{elliott1958growth,wood1982internal,townsend1965response,panofsky1973tower,panofsky1984atmospheric,pendergrass1984dispersion,savelyev2001notes,antonia1971response}], and the values previously inferred for White Sands IBL growth based on observations of dune dynamics \cite{jerolmack2012internal}. 
However, the downwind growth of $\delta_{IBL}$ in our simulations is not as smooth as simplified treatments would suggest (Fig. 2\textit{B}); there are fluctuations superimposed on the general trend.
In fact the second rise in $\delta_{IBL}$, that begins around $\hat{x} = 3000$ m, coincides with a subtle but persistent topographic rise underlying the dunes that was previously identified \cite{baitis2014definition, gunn2020macro}. 
Observed changes in flow across the roughness transition (Fig. 2\textit{A}) are not as abrupt as most models assume. 
The flow appears to respond continuously to changes in the spacing and geometry of dunes throughout the dune field.

\begin{figure*}[t!]
    \centering
    \includegraphics[width=1\linewidth]{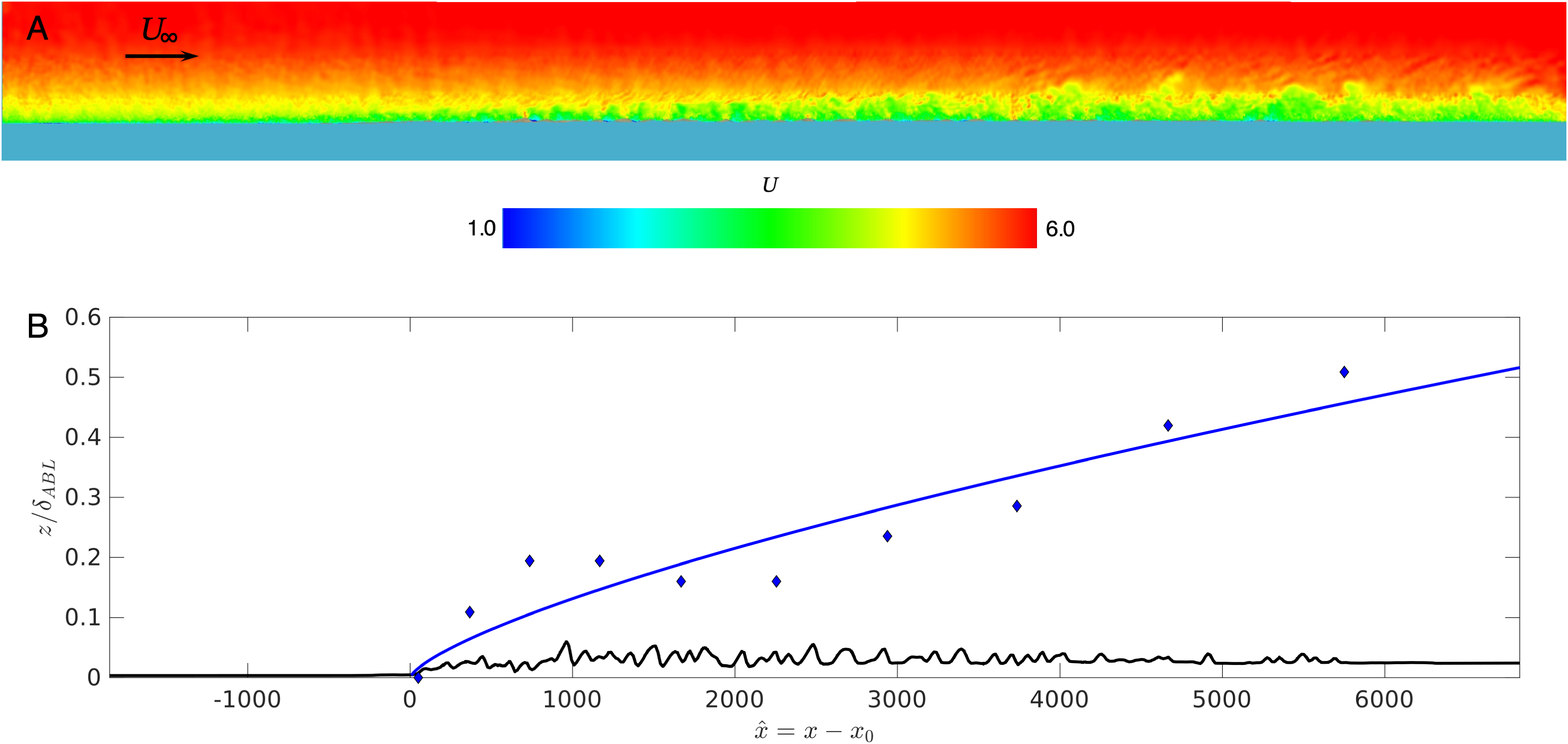}
    \caption{Development of the internal boundary layer. (\textit{A}) Instantaneous streamwise velocity in the spanwise center plane, flow going from left to right. (\textit{B}) Measured values of $\delta_{IBL}$ (black, blue-filled diamonds) with the power-law fit (blue line) given by $0.29x^{0.71}$, and a center-line profile of the dune field (black line).}
    \label{fig:ftwo}
\end{figure*}

\subsection*{Near-Wall Implications of the Roughness Transition}
\label{sec:tau}
Sand transport in dune fields is driven by the near-surface winds.
In particular, sediment transport equations relate sediment flux, $q_s$, to the local boundary stress, $\tau_b$, in excess of the entrainment threshold, $\tau_c$; commonly used equations have the form $q_s = K \sqrt{\tau_c}(\tau_b - \tau_c)$, where $K$ is a parameter related to sediment properties \cite{duran2011aeolian,gunn2020macro,kok2012physics,barchyn2014fundamental}. 
The presence of dunes is known to cause spatial variations in $\tau_b$ due to speedup and slow down of near-surface winds, which in fact drives the stoss-side erosion and lee-side deposition, respectively, that migrates dunes \cite{bagnold1941, weng1991air, kroy2002minimal, andreotti2002selection, livingstone2007geomorphology}. 
We examine spatial ($x$) variations in $\tau_b$ along the centerline of our model domain over the length of the modeled dune field (Fig. 3\textit{A}). 
The first-order observation is that topography and boundary stress co-vary as expected; dune crests are regions of high stress due to speedup, where $\tau_b$ is as much as four times as large as shielded troughs.
To examine any systematic downwind change in $\tau_b$ that results from IBL development, we must average the variations in stress over individual roughness elements (dunes). 
Here we perform a spanwise averaging over the model domain, $\Bar{\tau}_b$, in order to suppress the contribution of individual dunes and enhance the signal of the mescoscale IBL pattern (Fig. 3\textit{A}). 
Models predict that there should be a spike in $\tau_b$ at the location of a smooth $\rightarrow$ rough transition, followed by a gradual stress relaxation as the IBL develops downwind of the transition [\textit{SI Appendix}, Fig. S11~\cite{elliott1958thesis}]. 
The observed pattern in our simulations, however, is more complex and subdued than the idealized models.
Starting from the roughness transition ($\hat{x} = 0$), $\Bar{\tau}_b$ gradually increases downwind over the first $\sim 1$ km of the dune field. 
This may be because the dunes are superimposed on an underlying topographic ramp; i.e., the spanwise-averaged elevation rises significantly over the first 1 km of the dune field. 
In addition, dunes grow in size over the first 1 km of the dune field. 
Together, these factors likely drive an increase in boundary stress over this region. 
After $\hat{x} = 1$ km, $\Bar{\tau}_b$ slowly relaxes over several kilometers -- even though local stress peaks on individual dunes continue to be large. 
This gradual decline in $\Bar{\tau}_b$ must be the result of the developing IBL. 
This simulated pattern is consistent with the measured decline in time-averaged 10-m wind speed reported by Gunn \textit{et al.} \cite{gunn2020macro} from three meteorological towers along a transect at White Sands (their Figure 2). 
We cannot directly compare simulation results to sediment flux determined from dune migration. 
This is because our simulations use daytime-averaged wind conditions -- which produce boundary stress values that are less than the entrainment threshold -- whereas sand transport at White Sands only occurs (on average) for several hours per day during the peak windy season \cite{jerolmack2012internal, gunn2021circadian}. 
Nevertheless, the simulated reduction in $\tau_b$ due to IBL development is compatible with the observed decline in sand flux of about a half over the first $\sim 6$ km of the dune field \cite{jerolmack2012internal, gunn2020macro}. 

Even though simulated IBL growth roughly follows classic scaling behavior, the observed boundary stress pattern does not.
In particular, a common parameterization used for estimating $\tau_b$ in idealized IBL models over-predicts the stress response to roughness changes for White Sands [\textit{SI Appendix}, Fig. S11~\cite{elliott1958thesis}]].
This suggests that heterogeneity of roughness, and the sensitivity of turbulence to that roughness, produces a first-order departure from theory developed for idealized conditions. 
We look to the structure of the flow across the developing IBL -- in particular, the near-bed velocity {fluctuations} -- to better understand how the simulated downwind changes in boundary stress occur. 
We observe a systematic growth in the scale of coherent turbulent structures, that coincides with the growing IBL (Fig. 3\textit{B}). 
Qualitatively similar behavior is observed at other wall-normal elevations within the IBL (Fig. 3\textit{C}), {where an absence of turbulence exists outside of the IBL} suggesting that IBL growth may set the scale of growing turbulent structures within.

\begin{figure*}[tbh!]
    \centering
    \includegraphics[width=1\linewidth]{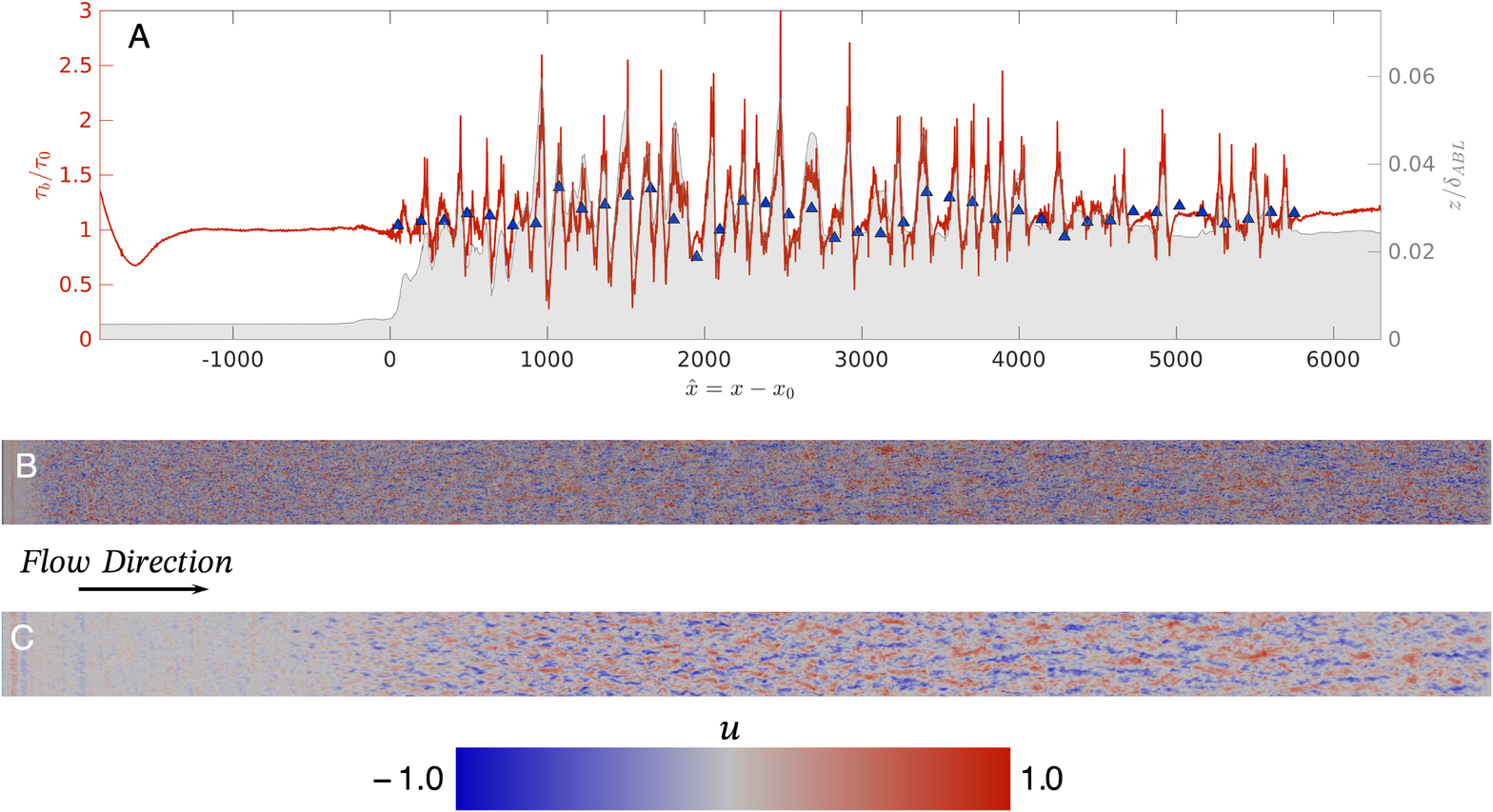}
    \caption{Changes to the near-boundary characteristics of the flow due to the development of the internal boundary layer. (\textit{A}) Evolution of the centerline time-averaged boundary stress normalized by the boundary stress upstream of the roughness transition (orange line) and the spanwise- and time-averaged boundary stress (black, blue-filled triangles). Values are overlaid on an outline of the centerline profile of the dune field (gray shaded area). (\textit{B}) Instantaneous streamwise velocity fluctuations at the first off-wall cell, projected onto the surface of the dune field. (\textit{C}) A wall-parallel plane at $z/\delta_{ABL} = 0.1$ showing instantaneous streamwise velocity fluctuations. 
    Flow is moving from left-to-right in (\textit{B,C}). 
    }
    \label{fig:fthree}
\end{figure*}

\begin{figure*}[tb!]
    \centering
    \includegraphics[width=1\linewidth]{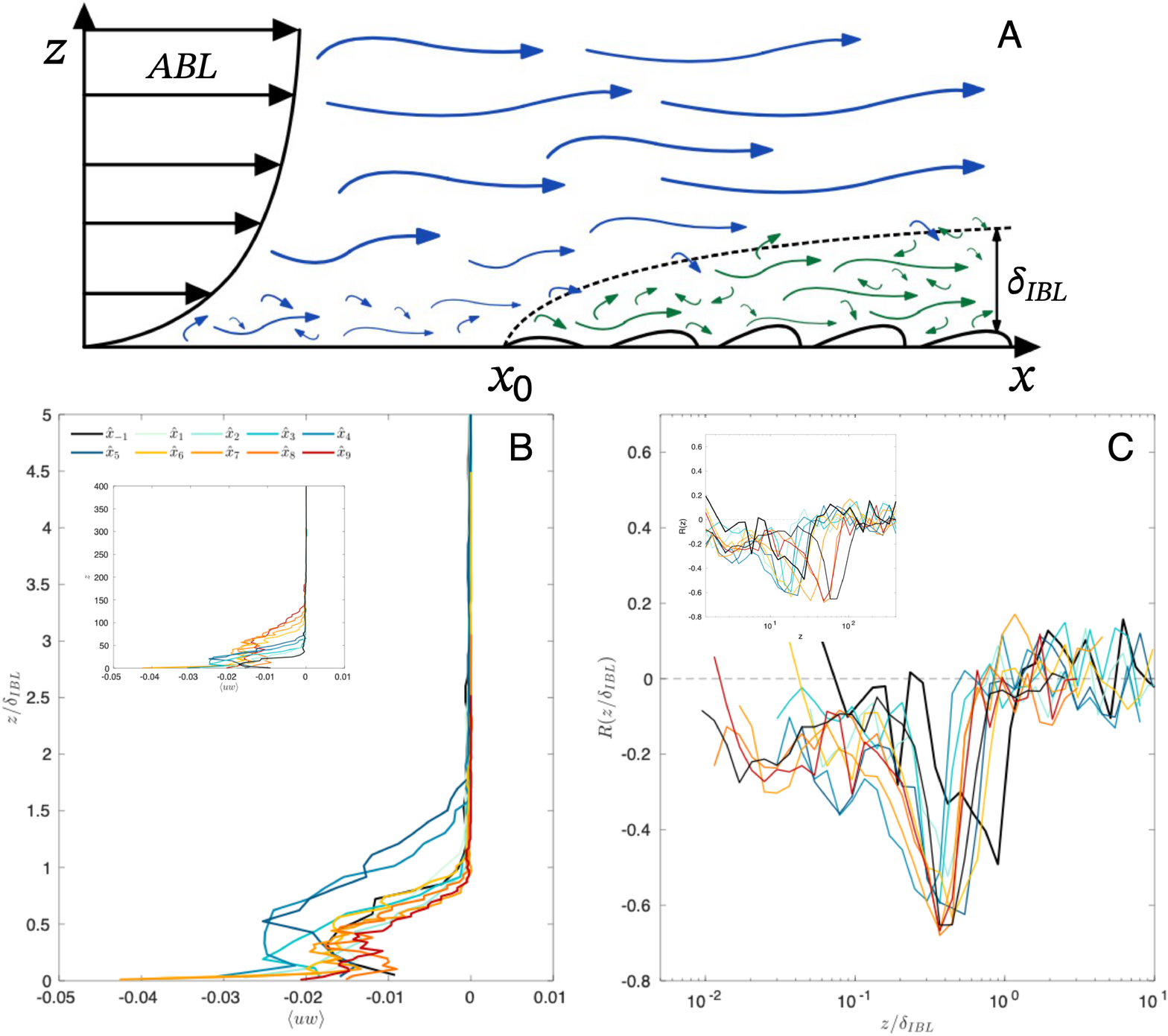}
    \caption{Self-similarity of turbulence within the IBL. (\textit{A}) Schematic of a smooth $\rightarrow$ rough transition. Flow is from left to right, and the internal boundary layer (dashed line) forms at the interface, $x_0$, delineating the transfer of momentum by the new surface and the outer flow region. The blue streamlines represent the turbulent flow within the ABL, and the green streamlines represent flow originating from within the IBL. The added turbulence further from the wall within the IBL is reflected by the thickening of $\langle uw\rangle$, and the negative $R_{AM}$ peak can be seen by the interactions between flow at the edge of the IBL and the ABL interface. (\textit{B}) The Reynolds shear stress through the dune field, with wall-normal location $z$ normalized by the relevant length scale $\delta$. For station $\hat{x}_{-1}$ in the upstream Alkali flat, and the first station $\hat{x}_1$ in the dune field, the length scale used is $\delta_{ASL} = 30$ m; for all downstream stations the length scale used is $\delta_{IBL}$ for each location, determined using methods described in the text. The first five locations within the dune field use a cool color scheme and the final four use a warm color scheme; associated distances from the roughness transition are shown with corresponding color gradients, light to dark. Inset shows the same data, not normalized. (\textit{C}) Profiles of amplitude modulation coefficients where $z$ is normalized in the same manner as \textit{B}; inset shows same data, not normalized. Colors and locations are the same as (\textit{B}).
    }
    \label{fig:ffour}
\end{figure*}

\subsection*{Self-Similarity of Turbulence within the IBL}
\label{sec:RSS_AM}


{The IBL acts as a mechanically distinct feature, which delineates the portion of the flow that retains a memory of the upstream wall-condition, and the part that adapts to the new condition. 
Previous work on a rough $\rightarrow$ smooth transition revealed that the IBL may 'shield' the outer-region of the flow, and that as the flow progresses downstream the energy contained outside the IBL is lost~\cite{hanson2016development}.
Through analysis of the Reynolds shear stress, $\langle uw\rangle$, a strong indicator of turbulence production, it becomes clearer just how much the IBL shields the outer-region from the momentum flux induced by the roughness (Fig. 4\textit{A}). 
Gul and Ganapathisubramani~\cite{gul2022experimental} showed that the IBL height $\delta_{IBL}$ corresponds to the location in the flow where the Reynolds shear stress diminishes to (near) zero for a smooth $\rightarrow$ rough tranistion (their Figure 1\textit{F}).
We examine the evolution of $\langle uw\rangle$, as a function of wall-normal position, on the Alkali Flat and at multiple stations downstream of the roughness transition.
As the flow progresses past $\hat{x} = 0$, we see that the region of elevated Reynolds shear stress thickens (Inset Fig. 4\textit{B}).
For the smooth Alkali flat, the region associated with elevated Reynold shear stress should be the Atmospheric Surface Layer, $\delta_{ASL}$, which is typically considered to be roughly 1/10 the thickness of the ABL \cite{huang2021investigating,zilitinkevich2002third,geernaert1988measurements}. Using our assumed $\delta_{ABL}$ = 300 m, we estimate $\delta_{ASL}$ = 30 m. This value is comparable to the estimated $\delta_{ASL} = 60$ m determined from observations in the well-studied desert of Western Utah -- an environment similar to White Sands. Normalizing $z$ by the relevant length scale $\delta$ -- $\delta_{ASL}$ on the Alkali Flat and the first dune station, and $\delta_{IBL}$ for each downwind station -- we find a decent collapse of the wall-normal Reynolds shear stress profiles (Fig. 4\textit{B}) and that $\langle uw\rangle$ approaches zero at roughly $z/\delta = 1$. 
Two downstream stations depart from the general collapse; we attribute this to the significant fluctuations in IBL height around the overall downstream trend. 
Nevertheless, the general pattern we observe is a self-similar Reynolds shear stress profile within the IBL, and that the height of the IBL corresponds to the location where turbulence production becomes negligible.}

Prior work has demonstrated that the large-scale (low-frequency) motions that exist in the outer-region of the boundary layer can influence the near-wall, small (high-frequency) scales ~\cite{mathis2009large,hutchins2007evidence}. 
This influence may be quantified using an Amplitude Modulation (AM) correlation coefficient, $R_{AM}$, following \cite{mathis2009large}: 

\begin{equation}
    R_{AM}(z) = \frac{ \langle u^+_L(z,t) E_L(u^+_s(z,t))\rangle }{ \sqrt{\langle{u^+_L(z,t)}^2\rangle} \sqrt{\langle{E_L(u^+_s(z,t))}^2\rangle} }.
    \label{R_AM}
\end{equation}

Here, $u_L^+$ is the large-scale component of the velocity fluctuations, $u_s^+$ is equivalently the small-scale component, and $E_L(u^+_s)$ represents the filtered envelope of the small-scale velocity fluctuations.
The process of amplitude modulation is outlined in Mathis \textit{et al.}~\cite{mathis2009large} and is included in more detail in (\textit{Materials and Methods}, \textit{SI Appendix}), but a brief overview is presented here. 
{A velocity signal, taken to be $u^+$, is decomposed into a large-scale and small-scale component using a spectral cutoff filter (\textit{Materials and Methods}, \textit{SI Appendix}, Fig. S12). 
Next, a Hilbert transformation is conducted on the small-scales to create an envelope of the signal, that is then subjected to an additional filtering step. 
The equivalent of} a Pearson coefficient~\cite{pope} is created using the large-scale signal and the filtered envelope of the small-scale signal to find $R_{AM}$ (\textit{Materials and Methods}).

We calculate a single-point $R_{AM}$ as a function of wall-normal distance at multiple streamwise locations, both preceding and following the roughness transition (Fig. 4\textit{C}). 
We first examine the vertical $R_{AM}$ profile over the smooth Alkali Flat. 
The most notable feature is the large negative correlation, which occurs at a wall-normal elevation of $z \approx 25$ m. 
Mathis \cite{mathis2009large} suggests that this is the result of intermittency arising in the outer region of the boundary layer, due to shear with the fluid above the Atmospheric Surface Layer (ASL).
Indeed, the $R_{AM}$ profile in our simulations is in good qualitative agreement with their experimental observations \cite{mathis2009large}.
They found that the negative peak occurs {between $z/\delta = 0.7$ and $z/\delta = 1.0$}.
{Using $\delta_{ASL} = 30$ m for our data suggests a relative height of $z/ \delta_{ASL} \approx 0.83$, within the range of the results of \cite{mathis2009large}.}
The same qualitative structure of the $R_{AM}$ is seen across the dune field. 
The magnitude of the prominent negative correlation is more or less preserved; however, its location, $z$, systematically shifts (Inset in Fig. 4\textit{C}). 
At the start of the roughness transition (near $\hat{x}$ = 0), the negative peak appears closest to the bed; moving downwind (increasing $\hat{x}$), the peak consistently shifts away from the bed toward higher elevations (Inset in Fig. 4\textit{C}). 
This suggests that the migration in this peak is set by the growing height of the IBL itself. 
We normalize the wall-normal height following the same procedure used for the Reynolds shear stress profiles, and find that the $R_{AM}$ profiles collapse onto a reasonably similar master curve. 
These results suggest that modulation of large-scale atmospheric turbulence within the IBL occurs in a self-similar manner, that scales with IBL height. 

\section*{Discussion}
For most studies examining Internal Boundary Layer development in response to changes in roughness, the idealized analytical solutions derived from classic scaling arguments are still the go-to model. 
While such closed-form solutions are convenient, they are inadequate for determining the near-bed turbulence and boundary stresses in Atmospheric Boundary Layer flows that are critically important for heat and water flux (evaporation), CO$_2$ (eddy covariance), dust emission and sediment transport. Our study shows how relaxing the assumption of a step change in roughness, and explicitly modeling natural heterogeneous topography, is essential for capturing the mesoscale flow behavior in the ABL that is of central importance for the evolution of landscapes and the activities of humans living within them. 
By resolving ABL turbulence using Wall-Modeled Large-Eddy Simulation, while carefully treating inlet/boundary effects and using a surface-conforming mesh for the topography, we were able to produce simulated flows that were validated against field lidar velocimetry data. 
Our results are consistent with what has been measured and inferred about IBL dynamics at White Sands from previous studies, while providing qualitative and quantitative insight on the mescoscopic feedback between flow and form that cannot be seen from field data alone.

While resolving large-scale fluid motions and heterogeneous boundary roughness is clearly important, our results also suggest that there may yet be some generic behaviors in the spatially growing IBL.
In particular, the self-similar profiles of Reynolds shear stress and amplitude modulation within the developing IBL indicate that, when present, the IBL is the relevant mesoscopic length scale governing turbulence in the near-surface flow. 
These findings suggest that if the IBL thickness is known, then aspects of the turbulence structure within it can be predicted. 
Qualitatively, the growing size of large-scale coherent flow structures downwind of the roughness transition coincides with the growing thickness of the IBL.
It is sensible that the mechanically distinct IBL somehow sets the scale for the largest eddies contained within it. This last point warrants further study. 

It is important to make the clear the limitations of our present study -- in their application to White Sands, and for the potential extrapolation to other settings. 
There are two important factors in wind dynamics that were neglected here. 
The first is buoyancy; the FATE campaign \cite{gunn2021circadian} showed that non-equilibrium buoyancy effects drive the sand-transporting winds at White Sands, and that typical frameworks like Monin-Obukhov Similarity Theory cannot account for the strength of convection effects on surface winds. 
The agreement of our simulation results with the time-averaged daytime winds from FATE indicates that (i) time averaging removes the buoyancy effect, and/or (ii) winds in the near-surface layer are sufficiently mixed by turbulence that buoyancy effects can be neglected -- at least in the lower 10s of meters \cite{huang2021investigating,zilitinkevich2002third,geernaert1988measurements}.
We were able to isolate the influence of roughness on driving relative changes in the boundary stress; this suggests that the roughness effect is, to first order, decoupled from the buoyancy effect. 
However, the magnitudes of our simulated stresses are lower than the sand-transporting winds at White Sands, which can only be resolved with transient and nonlinear buoyancy effects. 
The second neglected factor is non-stationarity of the flow. 
Winds in the ABL, including White Sands, are highly variable in magnitude and direction. 
Because desert sand dunes evolve over decades \cite{myrow_jerolmack2018}, a steady flow approximation is reasonable for examining feedbacks between dune roughness and the near-surface winds. 
This approximation may not be acceptable in other situations, however, where event-scale weather phenomena are of interest. 

Our study demonstrates a mesoscale coupling beteween flow and form that is relevant for landscape dynamics; the dunes alter the flow, while the flow pattern alters the dunes. 
The emergence of larger-scale structures indicates that modeling dunes in isolation, as is typically done, will not produce the correct stress profile. 
We suggest that a useful next step will be to examine how the presence of the IBL influences the wind stress profile over individual dunes. 
The mesoscopic interactions of ABL flows with heterogeneous roughness is also of central importance for cities, forests, and the ocean-land interface where wind may be carrying aerosols, dust or wildfire smoke \cite{kahn2008wildfire, li2017aerosol, miao2019interaction, baldocchi2020eddy, kok2023mineral}. 
The transport, deposition or bypass/ejection of these particulates from landscapes depends on the interaction between flows within the developing IBL and the flow outside of it. 
Recent improvements for modeling particles in LES \cite{park2017simple} could be introduced to simulations like ours, to track how the IBL modulates the transport of aerosols across and out of landscapes. 
Finally, the lidar velocimetry data that allowed validation of our model is rare. 
We suggest that carefully deployed field campaigns, coupled with well-resolved LES simulations, can allow researchers and practitioners to create 3D flow fields for many complex environmental flows.

\section*{Materials and Methods}

\subsection*{Details of WMLES}
In this study, we use \textit{Charles}, from Cascade Technologies (Cadence Design Systems), which is an unstructured grid, body-fitted finite-volume LES flow solver. 
\textit{Charles} solves the compressible Navier-Stokes equations in a low-Mach isentropic formulation, using a second-order central discretization in space, and a second-order implicit time-advancement scheme~\cite{gorle2023investigation}. 
The solver has been deployed for many high Reynolds number turbulent flow cases, including LES over the Japanese Exploration Agency Standard Model~\cite{goc2021large} and atmospheric boundary layer flows over buildings~\cite{gorle2023investigation,hwang2022large}, as well as wall-bounded flows with roughness~\cite{cooke2023numerical}. 
The entire code is written in C++ and deploys Message Passing Interface for parallelization. 
Part of this work used Anvil at Purdue University through allocation MCH230027 from the Advanced Cyberinfrastructure Coordination Ecosystem: Services \& Support (ACCESS) program, which is supported by National Science Foundation grants \#2138259, \#2138286, \#2138307, \#2137603, and \#2138296~\cite{access}.

\textit{Charles} uses an isotropic Voronoi meshing scheme which allows for highly accurate body-fitted meshes suitable for complex, irregular geometries. Mesh design requires an outer, far-field grid spacing, $\Delta_{FF}$, in which subsequent refinement levels are built off. This is based on $\Delta_{FF}/2^n$, where $n$ is the desired number of refinement levels~\cite{cooke2023numerical}. More details on the meshing technique may be found in Lozano-Dur\'an, Bose, and Moin~\cite{lozanoduran2022performance}. 

The grid used for the study contained approximately $85\times10^6$ control volumes, with $\Delta_{FF} = 28$ m and a finest cell-spacing of $\Delta_{min} = 0.75$ m. In viscous units, the near-wall spacing is $\Delta^+_{min} \approx 3200$, due to the high $Re_\tau$, and is the same for the streamwise, spanwise, and wall-normal grid spacing due to isotropy of the cells. We adequately resolve the height of the ABL with nearly 67 control volumes, and the IBL using between approximately 20 control volumes near the roughness transition, to 46 control volumes at the end of the dune field. A refinement study of the mesh was conducted to ensure convergence of quantities of interest (\textit{SI Appendix}, Figs. S6 and S7).    

The domain is sized to be $8.6\times0.5\times1$ $km^3$, using a domain study to ensure the spanwise width of the domain had no effect on the flow, due to the symmetry boundary conditions placed on the side walls (\textit{SI Appendix}, Figs. S3-S5). 
The top wall of the domain also deploys a symmetry boundary condition, and we observe no influence on the flow. 
The inflow and outflow regions were artificially extended to allow for development of the incoming turbulent boundary layer and the placement of a numerical sponge outlet condition. 
A sponge region is used at the outflow to prevent pressure waves from reflecting back into the domain and causing numerical instabilities~\cite{bodony2006analysis,mani2012analysis}. 
The sponge region is not considered in the analysis. 
To reduce computational cost of the simulation, an algebraic wall-model boundary condition~\cite{bose2018wall} is imposed on the topography, where the matching-height for the wall-model is placed at the center of the first cell.
More information on the inflow method to generate turbulence is provided in (\textit{SI Appendix}).

\subsection*{Amplitude Modulation}

For the calculation of $R_{AM}$, we probe for $U$ at the same logarithmically-spaced streamwise stations within the dune field, at multiple wall-normal locations. 
A fluctuating signal, $u^+$, is then found at each probe point, and filtered. 
To conduct the filtering technique, a spectral cutoff filter is deployed.
This filter uses a cutoff wavelength $\lambda_{x,c} = \delta_{ABL}$, where $\lambda_x \equiv U_c/\omega$ is recovered using Taylor's hypothesis, selecting the mean velocity at each wall-normal position as the convective velocity~\cite{anderson2016amplitude}.
The signal is transformed into the frequency domain using the Fourier transform, and the cutoff filter is applied at this stage to get the large-scale features of the flow, $\hat{u}^+_L$. 
The signal is transformed back to the physical space, and the small-scale features are found by subtracting the filtered signal from the raw signal, $u^+_s = u^+ - u^+_L$. 
Next, a Hilbert transformation is conducted on the small-scale signal.
This transformed signal is again filtered using the technique described above, and we are left with the filtered envelope of the small-scale signal.
The process is repeated for every wall-normal location at every streamwise station.

To find $R_{AM}$ and complete the process described above, long time-series data are required. 
Generally, for $R_{AM}$ to be converged, it is recommended the flow experience $5,000 \leq TU_\infty/\delta \leq 14,000$~\cite{mathis2009large,anderson2016amplitude}, where $TU_\infty/\delta$ is a non-dimensional large-eddy turnover time. 
{For our calculation, we define $\delta \equiv \delta_{ABL}$.}
Due to the stringent computational cost of this analysis, it is unfeasible to conduct the minimum suggested turnover times, so only $\sim 1000TU_\infty/\delta$ were completed for this analysis. 
We regard this number as a reasonable time for convergence of $R_{AM}$, especially in the context of ABL flows, and we show a lower threshold of $TU_\infty/\delta$ may be allowable (\textit{SI Appendix}, Fig. S13). 
Given this, the results presented are a means to represent what might be expected of AM at mesoscopic scales.

\begin{acknowledgments}
We would like to acknowledge Prof. Andrew Gunn for helpful discussions related to his work at White Sands, and for providing his experimental data. 
G.P and J.C. acknowledge the support from the University of Pennsylvania (faculty startup grant and the Fontaine fellowship) and the National GEM Consortium Fellowship.
D.J.J. was supported by NASA PSTAR (Award 80NSSC22K1313).
\end{acknowledgments}

\newpage

\bibliography{dunes.bib}

\end{document}